\documentclass[nofootinbib,aps,pra,twocolumn,english]{revtex4-1}

\pdfoutput=1

\usepackage{orcidlink}

\usepackage{graphicx}
\usepackage{wrapfig,enumerate,slashed}
\usepackage[utf8]{inputenc}
\usepackage{adjustbox}
\usepackage{color, colortbl}
\usepackage{tabularray}

\usepackage[english]{babel}
\usepackage[T1]{fontenc}
\usepackage{amsmath}

\usepackage{float}

\usepackage{appendix}
\usepackage{placeins}

\usepackage{amsfonts}
\usepackage{amsmath}
\usepackage{wasysym} 
\usepackage{graphicx}
\usepackage{comment}
\usepackage{slashed}
\usepackage{booktabs}
\usepackage{listings}
\newcommand{\be}{\begin{eqnarray}}
\newcommand{\ee}{\end{eqnarray}}

\newcommand{\bee}{\begin{eqnarray}}
\newcommand{\eee}{\end{eqnarray}}
\newcommand{\beeq}{\begin{equation}}
\newcommand{\eeeq}{\end{equation}}

\usepackage{listings}
\usepackage{color}
\usepackage{hyperref}
\usepackage{tabularray}

\usepackage{dsfont}

\newcommand{\cc}[2]{c\genfrac{[}{]}{0pt}{}{#1}{#2}}

\def\ie{{\it i.e.}\/}
\def\eg{{\it e.g.}\/}

\def\aka{{\it a.k.a.}\/}

\definecolor{dkgreen}{rgb}{0,0.6,0}
\definecolor{gray}{rgb}{0.5,0.5,0.5}
\definecolor{mauve}{rgb}{0.58,0,0.82}
\definecolor{cyan}{rgb}{0.88,1,1}
\definecolor{TopRow}{rgb}{0.4,0.7,1}
\definecolor{lblue}{rgb}{0.8,0.9,1}

\makeatletter

    \def\CT@@do@color{%
      \global\let\CT@do@color\relax
            \@tempdima\wd\z@
            \advance\@tempdima\@tempdimb
            \advance\@tempdima\@tempdimc
    \advance\@tempdimb\tabcolsep
    \advance\@tempdimc\tabcolsep
    \advance\@tempdima2\tabcolsep
            \kern-\@tempdimb
            \leaders\vrule
                    \hskip\@tempdima\@plus  1fill
            \kern-\@tempdimc
            \hskip-\wd\z@ \@plus -1fill }
    \makeatother
\begin{document}
\title{String Model Building on Quantum Annealers}

{{\hfill CERN-TH-2023-123, IPPP/23/29}\vspace{0.5cm} }

\author{Steven~Abel \orcidlink{0000-0003-1213-907X}$^{1,2,3}$ }
\email{steve.abel@durham.ac.uk }
\author{Luca~A.~Nutricati \orcidlink{0000-0002-5045-5113}$^{2,3}$}
\email{luca.a.nutricati@durham.ac.uk}
\author{John Rizos \orcidlink{0000-0001-7586-1700}$^{4}$}
\email{irizos@uoi.gr}

\affiliation{\vspace{0.3cm} $^{1}$Theoretical Physics Department, Cern, 1211 Geneva 23, Switzerland}
\affiliation{\vspace{0.0cm}  $^{2}$Institute for Particle Physics Phenomenology, Durham University, Durham DH1 3LE, UK}
\affiliation{$^{3}$Department of Mathematical Sciences, Durham University, Durham DH1 3LE, UK}
\affiliation{$^{4}$Department of Physics, University of Ioannina
GR45110 Ioannina, Greece}

\begin{abstract}
{\small
We explore for the first time the direct construction of string models on quantum annealers, and investigate their efficiency and effectiveness in the model discovery process. Through a thorough comparison with traditional methods such as simulated annealing, random scans, and genetic algorithms, we highlight the potential advantages offered by quantum annealers, which in this study promised to be roughly fifty times faster than random scans and genetic algorithm and   approximately four times faster than simulated annealing. 
}
\end{abstract}

\maketitle
  \tableofcontents

% make two column:

\flushbottom

%%%%%%%%%%%%%%%%%%%%%%%%%%%%%%%%%%%%%%%%%%%%%%%

\section{\label{Sec:Intro}Introduction}

There is continued interest in the problem of model selection in string theory which, due to the vast number of models available, presents a fascinating ``big-data'' challenge. Indeed depending on the set-up, the various estimates of the number of available models in the parameter space vary from the original $10^{500}$ estimate in type IIB flux compactifications \cite{ Douglas:2003um, Ashok:2003gk} to significantly larger numbers, for example $10^{272,000}$ F-theory flux compactifications on a single elliptically fibered four-fold~\cite{Taylor:2015xtz}. In fact the number of Standard Model (SM) -like compactifications could itself be as large as $10^{700}$ \cite{Constantin:2018xkj}. 

Given this challenge attention has naturally turned to heuristic search methods. The argument for pursuing them is that nature itself finds solutions to problems without any difficulty within similarly large (or indeed much larger) search spaces. Heuristic methods typically attempt to mimic these natural processes. In the string context perhaps the most successful heuristic approach to date has proven to be the genetic algorithm (GA) \cite{holland1975adaptation,goldberg1989genetic}, which mimics evolution by implementing selection and breeding cycles on a population of would-be string solutions. Such methods have been shown capable of  achieving orders of magnitude speed-up over blind scans in a string setting
 ~\cite{Blaback:2013ht,Abel:2014xta,Ruehle:2017mzq,Cole:2019enn,AbdusSalam:2020ywo,Ruehle:2020jrk,Bena:2020xrh,Larfors:2020ugo,Bena:2021wyr,Abel:2021rrj,Cole:2021nnt,Loges:2021hvn}.

In this paper we shall consider a different class of heuristic search method in the string context, namely quantum adiabatic algorithms \cite{farhi2000quantum,das2008quantum}. Here we focus on the particular implementation of adiabatic computing known as Quantum Annealing (QA), in which the problem to be solved is mapped to to the minimisation of an  Ising Hamiltonian. The full Hamiltonian in quantum annealing comprises an admixture of this Ising  {\it problem-Hamiltonian} and a trivial Hamiltonian for which the ground state is known. The original idea behind quantum annealing (and quantum adiabatic algorithms more generally) is to begin in the ground state of the trivial system and adiabatically replace the trivial Hamiltonian with the problem Hamiltonian, while remaining in the ground state throughout. Provided we can remain in the ground state the final configuration will yield a solution to the problem. More modern approaches have extended this idea (for example using reverse annealing, of which more later) however the basic principle of arranging an interplay between a problem Hamiltonian and a trivial Hamiltonian is universal. 

Quantum annealing has been utilised in many simple settings, notably for solving network problems, but its application in high energy theory has up to now been somewhat limited  (see Refs.~\cite{Mott:2017xdb, Abel:2020ebj, Abel:2020qzm,ARahman:2021ktn,Schenk:2020lea,   ARahman:2021ktn,Abel:2021fpn, Pomponio:2021ltz,  Lagnese:2021grb, Bando:2021opl,Illa:2022jqb,Criado:2022aoo, Fromm:2022vaj,Abel:2022bln,Abel:2023zwg,Kim:2023sie} for some examples). 
In the string setting, quantum annealing has been employed in hybrid algorithms, for example most recently in Genetic Quantum Annealing \cite{Abel:2022bln,Abel:2023zwg} in which the GA performance is improved with a quantum annealing stage, however it has not to date been possible to implement a full string search directly on a quantum annealer (or indeed on a quantum computer of any kind). Nevertheless the  recent study in Ref.~\cite{Abel:2022wnt} laid the foundations for achieving this goal by showing that quantum annealers can solve the kind of discrete problems (for example satisfying anomaly cancellation conditions) that one typically encounters in string model building. 
% QNN  Abel:2022lqr,}

In the present study we will break new ground by for the first time implementing string models directly on a quantum annealer, and using the annealer to search for string theories with SM-like properties. 
The general approach we shall use is a combined technique which embeds the string consistency conditions (\aka~the GSO conditions) themselves on the annealer, but which performs  certain additional phenomenological checks (such as for example selecting only models with three generations) during a second step. In such an arrangement the quantum annealer is essentially providing a consistency filtering of models, which may then be classically tested against other constraints. We shall see that this approach significantly enhances the overall efficiency of the algorithm. It can  successfully be used to search parameter spaces orders of magnitude more quickly than either a blind scan {\it or} more traditional classical heuristic methods such as GAs.

\section{\label{Sec:Annealer}Encoding string models on a quantum annealer}

We begin with the formulation of the models themselves. As described this is done using an Ising Hamiltonian: hence our first and indeed most difficult task is to encode the consistency conditions of the string theory by reformulating their solution as the minimisation of a function $H(\sigma_\ell)$ of spin variables $\sigma_\ell ~=~ \pm 1$, where $\ell$ labels the spin sites. This  is the aforementioned {\it problem-Hamiltonian}, which is quadratic in the spins,
\begin{equation}
\label{eq:isingH}
 H(\sigma_\ell) ~=~ \sum_\ell h_\ell \sigma_\ell + \sum_{\ell m} J_{\ell m} \sigma_\ell \sigma_m~,
\end{equation}
where each $\sigma_\ell$ corresponds to physical $Z$-spins on the machine.

The particular models we shall consider are the $SO(10)$ models described in Ref.~\cite{Faraggi:2003yd}. The consistency conditions for these models amount to a set of generalized GSO (GGSO) projections determined by phases. The essential ingredients are described in Appendix~\ref{app1}, to which we will continue to refer. 
The important aspect of these consistency conditions for the present study is that they  can be written as a set of single qubit binary equations $f_A(\tau_i)=0$, where $\tau_{i} \in \{0,1\}$
map directly to the  GGSO phases of $0$ or $\pi$ which determine the particular string model. 
The binary $\tau$ variables can in turn be mapped to annealer spins as
\begin{equation}
\label{eq:taufromsig}
    \tau_{i} ~=~\frac{1}{2} (1+\sigma_{i})~.
\end{equation}
Thus these are arguably the string models that can most readily be  encoded in an Ising spin model.
As we are solving equations of purely single digit binaries we will use the names of the variables themselves to stand for the binary qubit value. 

To get a broad idea of the form of the GGSO constraints, they partially consist of six (three for spinorial and three for vectorial representations) systems of four linear equations each, which can be synthesised in the following expression:
\begin{equation}
\Delta^I U_i^I ~=~ Y_i^I  ~~ {\rm mod}~2 \, ,~~~~{\rm with}~ I=1,2,3 ~ {\rm and}~ i = s,v \, ,
\label{eq:constraints_1}
\end{equation}
where each choice of $(i,I)$ corresponds to a linear system of four equations (specifically $i = s,v$ refers to spinorial/vectorial representations and $I=1,2,3$ refers to the three orbifold planes, respectively. See Appendix~\ref{app1} for more details). The GGSO coefficients appear as components of the $Y_i^I$ vectors (defined in Eqs.~\eqref{eq:Y_s}, \eqref{eq:Y_v}) and also as entries in the $\Delta^I$ matrices defined in Eqs.~\ref{eq:delta_1}, \ref{eq:delta_2}, \ref{eq:delta_3}. Being related to phases, these coefficients take values of $0$ and $1$. Finally, $U_i^I$ is a vector of four elements which denote solutions of the $(i,I)$ system, also with entries in $\{ 0,1 \}$. We shall refer to
  the set of the solutions of the system as $\Xi_i^I$: it can contain at most $2^4 = 16$ solutions due to the binary nature of the components of the $U_i^I$ vectors.

Once Eq.~\eqref{eq:constraints_1} is satisfied, \ie, once we have a consistent GGSO projection, we will as described in the introduction further demand that a viable model must have $3$ generations by imposing the following constraint (in which $i\equiv s$) classically  \cite{Faraggi:2004rq}:
\begin{equation}
    N_F ~=~ \sum_{I=1}^{3} \sum_{p,q,r,s \in \Xi^I_s} X^{(I)}_{pqrs} ~=~ 3\, ,
\label{eq:generations_constraints}
\end{equation}
where $X^{(I)}_{pqrs} = \exp(i \pi \chi^{(I)}_{pqrs})$ and $\chi^{(I)}_{pqrs}$ are defined in Eqs.~\eqref{eq:chi^1}, \eqref{eq:chi^2} and \eqref{eq:chi^3}, for $I=1,2,3$, respectively. We shall also require two additional properties: existence of at least one SM Higgs doublet and existence of a top Yukawa coupling. The first requirement corresponds to having at least one solution coming from one of the $\{ (v,I), \, I =1,2,3 \}$ systems. The second requirement will be guaranteed by a particular choice of $U_v^3$ as we shall see in the following. 

Having given an overview of the fundamental ingredients and phenomenological requirements that we will impose, let us now describe our method in detail. To encode the GGSO constraints we adopt a technique that is significantly different from those that have been used to analyse this specific class of models before,  \eg  ~in Refs.~\cite{Faraggi:2003yd, Faraggi:2004rq, Faraggi:2006bc, Faraggi:2006pk, Faraggi:2007ms}. Indeed the $U_i^I$ parameters are typically scanned over along with the other variables corresponding to the GGSO phases. Here by contrast we first fix the values of $U_s^I$ and $U_v^I$ on the three orbifold planes which allows us to then search for suitable values of the GGSO coefficients. That is, following the discussion in Appendix~\ref{app1}, we can first without loss of generality set
\begin{equation}
U_s^1 = U_s^2 = U_v^3 =
     \left(
     \begin{array}{c}
        0\\
        0\\
        0\\
        0
\end{array}   
    \right) \, .
\label{eq:simp}
\end{equation}
Note that we fix $U_v^3$ to zero, which guarantees the existence of a top Yukawa coupling as we shall see. However we are still free to fix the residual parameters in $U_i^I$. For this study it is convenient to compare the different methods by studying the models with
\begin{align}
U^3_s=
\left(
\begin{array}{c}
0\\
1\\
0\\
0
\end{array}
\right)
\ ,\ 
U^1_v=
\left(
\begin{array}{c}
1\\
0\\
0\\
0
\end{array}
\right)
\ ,\ 
U^2_v=
\left(
\begin{array}{c}
1\\
1\\
1\\
1
\end{array}
\right)\, .
\label{eq:U_choice}
\end{align}
Of course in a full treatment one would scan through the $2^{12}\approx 4000$ possible choices of $U_s^3$, $U_v^1$, $U_v^2$. Having fixed these parameters, the remainder of the  GGSO constraints may be solved by quantum annealing. That is we are required to encode and solve the following equations on the quantum annealer:
\begin{equation}
    \begin{cases}
        \Delta^3 U_s^3 ~=~ Y_s^3 \\
        \Delta^1 U_v^1 ~=~ Y_v^1 \\
        \Delta^2 U_v^2 ~=~ Y_v^2 \\
        \chi_{pqrs}^{(3)} ~=~ 0        
\end{cases} ~~~~ \text{mod}~2 \, ,
\label{eq:system_annealer}
\end{equation}
where all the quantities involved can be expressed in terms of the GGSO coefficients following the definitions in Appendix~\ref{app1}. Comparing Eqs.~\eqref{eq:system_annealer} with Eq.~\eqref{eq:constraints_1}, one may wonder why spinorial projectors appear only on the third plane along with the corresponding chirality constraint. A similar question applies to the vectorial projectors, which are only present on the first and second plane. Indeed, it is straightforward to show that the choice in Eq.~\eqref{eq:simp} (see Appendix~\ref{app1} for details) trivialises the corresponding constraints on the first and second planes for spinorials as well as those on the third plane for vectorials. In a similar fashion, the chirality constraints on the first and second planes are satisfied by the conventions adopted in Eq.~\eqref{eq:a29}. 

The first three lines in Eq.~\eqref{eq:system_annealer} correspond to four linear equations, yielding a total of $13$ equations with $27$ unknown GGSO coefficients. Thus, we take  the corresponding problem Hamiltonian to be effectively a ``loss-function'' for this set of equations, which is to say that we take it to be the sum of the squares of the $13$ equations with additional integer parameters $K_{i=1,...,13}$ to absorb the modulo $2$ operation,
\begin{align}
    H ~=~ &\sum_{i=1}^{4} \left(\Delta^3_{ij} U_{s,j}^3 - Y_{s,i}^3 - 2K_i \right)^2 \nonumber \\
    & ~~~~~ + \sum_{i=1}^{4}  (\Delta^1_{ij} U_{v,j}^1 - Y_{v,i}^1 - 2K_{4+i})^2  \nonumber \\
    & ~~~~~~~~~ + \sum_{i=1}^{4}  (\Delta^2_{ij} U_{v,j}^2 - Y_{v,j}^2 - 2K_{8+i})^2 \nonumber \\
    & ~~~~~~~~~~~~~~~ + \left( \chi_{pqrs}^{(3)} - 2K_{13} \right)^2 \, ,
    \label{eq:annealer_hamiltonian}
\end{align}
where the sum over $j$ in each square is to be understood. The auxiliary variables $K_{i=1,...,13}$ are encoded using binary representations and take values in $[-3,4]$, while $\Delta$ and $Y$ are binary variables in $\{0,1\}$, which are straightforwardly encoded in annealer spins via Eq.~\eqref{eq:taufromsig}. This means that even for this restricted choice of $U^I_i$ the parameter space is $2^{28}\approx 10^8$. 

%%%%%%%%

Once models have  been acquired from this quantum annealing stage, they as mentioned need to be post-filtered classically to satisfy our additional phenomenological requirements. As discussed these conditions include the imposition of three generations, and the requirement of at least one Higgs doublet. However as explained in Appendix~\ref{app1} the third constraint, namely the existence of a top Yukawa coupling, is already ensured by our choice of $U_\nu^3$ and by  Eqs.~\eqref{eq:yukawa_1} and~\eqref{eq:yukawa_2} and is therefore already encoded in Eq.~\eqref{eq:annealer_hamiltonian}. Thus, from on now on we need focus only on the first two conditions.

Let us start by analysing the 3 generations constraint. As already mentioned, the chirality on the first and on the second planes is set to one thanks to the conventions adopted in Eq.~\eqref{eq:a29}. Therefore, the only chirality that needs to be checked is that on the third plane, which translates into the following equation (also without loss of generality):
\begin{equation}
    \sum_{p,q,r,s \in \Xi_s^3} X^{(3)}_{pqrs}  ~=~ \sum_{p,q,r,s \in \Xi_s^3} e^{i \pi \chi^{(3)}_{pqrs}} ~=~ 1\, .
    \label{eq:chi0}
\end{equation}
As the reader may have already noticed, to ensure the desired number of generations we must sum over {\it all} the spinorial solutions on the third plane. However, it may happen that the proposed solution $U_s^3$ in Eq.~\eqref{eq:U_choice} is not unique. In other words, since we are not imposing any constraint on the number of solutions, nothing prevents the system representing the spinorial projectors on the third plane from having solutions in addition to that already designated in Eq.~\eqref{eq:U_choice}. Indeed, the Hamiltonian in Eq.~\eqref{eq:annealer_hamiltonian} does not contain a term that enforces the uniqueness of that solution: it simply guarantees that $U_s^3$ is a solution. Therefore, in order to ensure the fulfillment of the generations constraint, we post-process all the candidate models proposed by the annealer and discard those that have additional solutions besides the one that we have already assigned. Similar arguments hold for the number of vectorials, however in this case the constraints are less severe as we only require there to be at least one Higgs doublet, that is we require at least one solution coming from one of the planes.

\section{Results and performance comparison with other search methods} 

In order to perform our analysis the system in Eq.~\eqref{eq:annealer_hamiltonian} was implemented  on D-Wave's \texttt{Advantage\_system4.1} architecture ~\cite{LantingAQC2017}: this annealer contains 5627 qubits, connected in a \emph{Pegasus} structure, but only has a total of 40279 couplings between them. For more details of the physical realisation of the system the reader is referred to Ref.~\cite{Abel:2022wnt}\footnote{It is worth mentioning that the quadratic Hamiltonian restriction does not apply to implementations of adiabatic quantum algorithms on gate quantum computers (at the expense of gate depth), as has been implemented in the Qibo architecture \cite{qibo_paper}. We return to this point later.}. 

The crucial aspect of these systems for our discussion is the anneal process itself. The entire Hamiltonian on the annealer takes the form 
\be 
{\cal H}(t) ~=~ A(t) H(\sigma_\ell^Z ) ~+~ B(t) \sum_\ell \sigma_\ell^X~,
\ee
where the overall couplings $A(t)$ and $B(t)$ are adjusted during the anneal, and where the second piece, $\sum_\ell \sigma _\ell ^X$ is the trivial Hamiltonian. 

The adiabatic annealing paradigm chooses $A(t_{\rm init}) = 0$ and $A(t_{\rm final}) = 1$ while $B(t_{\rm init}) = 1$ and $B(t_{\rm final}) = 0$.
However the quantum annealing method we use consists of several iterations of reverse annealing. Reverse annealing is a variation of the traditional quantum annealing approach: instead of starting the anneal in the groundstate of the trivial Hamiltonian, reverse annealing allows one to begin with the Hamiltonian as pure problem Hamiltonian ${\cal H}(t_{\rm init}) = H(\sigma_\ell ^Z)$, with the qubits initialised in a specific eigenstate (most likely not its ground state). One then adjusts the system to parametrically approach the trivial Hamiltonian which induces a controllable ``hopping'' of $\sigma^Z_\ell$ spins before then returning to the pure problem Hamiltonian. In other words in a single anneal iteration we take 
$A(t_{\rm init}) = A(t_{\rm final}) = 1$ and $B(t_{\rm init}) =B(t_{\rm final})= 0$, with non-zero values in between. The spins are then read from the annealer and used to initialise the next iteration. 

Repeating the process and  initialising each time with the best solution (in a fashion reminiscent of {\it elitism} in a GA), induces a kind of {\it quantum gradient descent} towards a solution over several iterations. This can in general lead to faster convergence and improved solutions.
\begin{figure}[h!]
\centering
\includegraphics[keepaspectratio, width=0.50\textwidth]{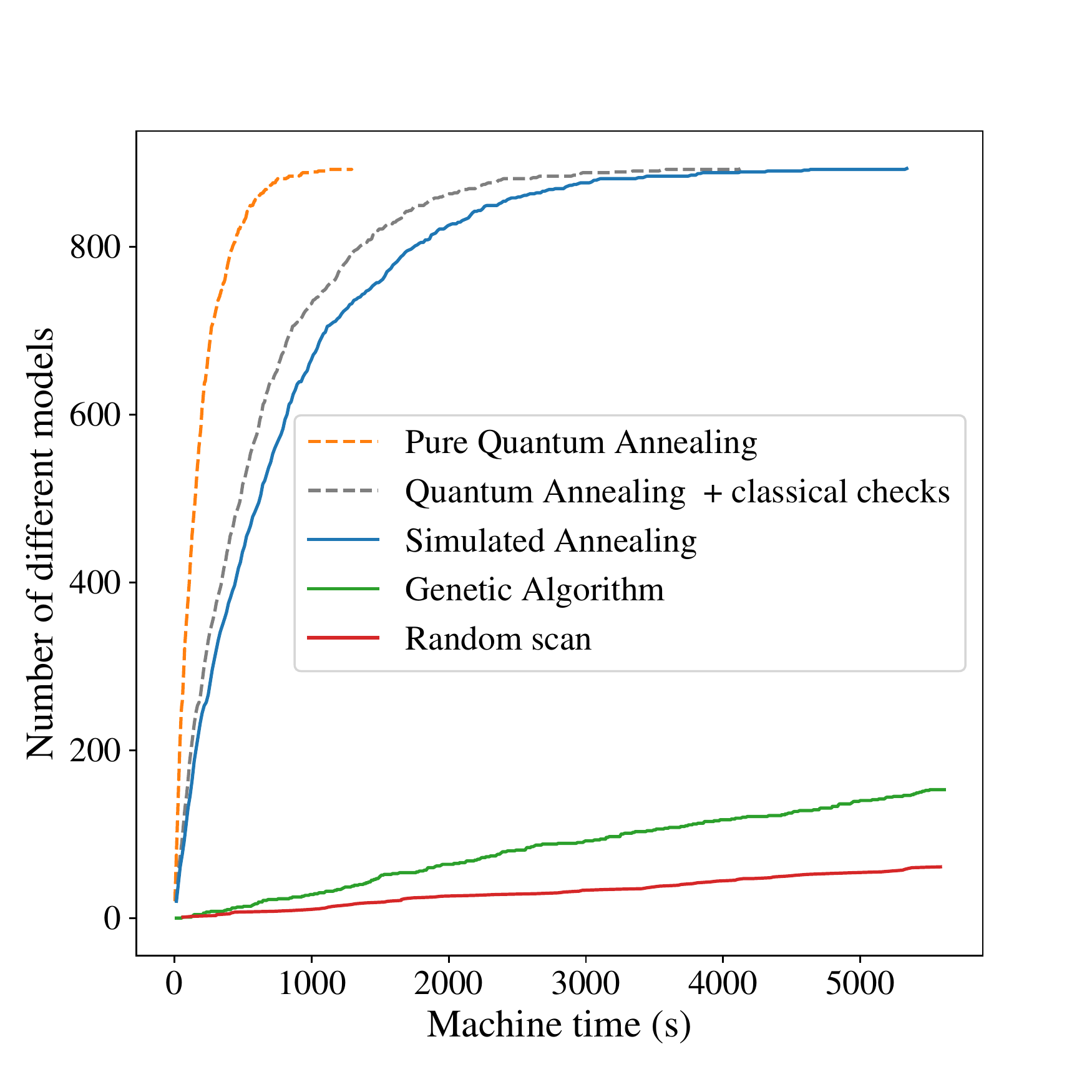}
\caption{ 
Comparison of the machine-time efficiency of various techniques for finding viable models. The methods analysed are: random scans, genetic algorithms, simulated annealing and quantum annealing. For the first three methods ``machine-time'' is equivalent to CPU-time. In the latter case it implies time on the annealer. The number of independent GGSO coefficients is 28, yielding to a search space of $ 2^{28} \sim 10^8$ possible models. The superiority of the methods using annealers is clearly evident in this case, surpassing the other techniques. The total number of models found using simulated annealing is 894 in approximately $6000s$ {\it vs} 153 and 61 models found using a genetic algorithm and a random scan, respectively. For the quantum annealer each anneal run had 2000 reads and used auxiliary variables $K_i \in [-3,4]$. The dashed lines both indicate quantum annealing and are projections based on the simulated annealing data. In these two cases the machine-time does not take into account the dead-time required due to bottlenecks in exchanging information with the annealer (which incurs a delay of order a second, and which could be removed if a portion of the machine were dedicated to the search). The machine time is computed based on the anneal schedule time only, which is fixed to $160 \,\mu s$. The grey line takes also into account the time required to do all the classical checks on the models found by the annealer. The orange line is an estimation of the performance of a quantum annealer supposing that all the classical checks are also encoded in the annealer Hamiltonian. } 
\label{fig:sat_plot_2}
\end{figure}

To assess the different methods we first note that a comprehensive scan of this system is possible,  and this was performed for comparison: this took about $48$ hours on a DELL PowerEdge R630 workstation with
32 GB of memory which resulted in a total number of $1024$ models, \ie, the search space of the set of models we have described contains one viable SM-like model in $10^5$. The second method we considered was simulated annealing optimised with a linear $\beta$-schedule. We also for comparison implemented the same system in a genetic algorithm. The GA was implemented with a fitness function defined linearly by the ranking, and optimised to find the best values for the mutation rate and for the learning rate (\ie, the number of times breeding occurs for the fittest individual compared to the least fit): these were found to be $3\%$ and $4$, respectively. 

To compare these four methods, we examined the rate at which each finds acceptable models in terms of machine-time. Apart from quantum annealing, all the other algorithms were run on standard computers. The results are collected in Fig.~\ref{fig:sat_plot_2} where we plot number of found SM-like theories against machine-time.  The definition of machine-time on quantum annealers is complicated by the fact that currently they are shared resources, and need to handle the influx of traffic and requests from multiple users simultaneously. This can lead to bottlenecks, resulting in longer wait times of a few seconds for accessing the machine or retrieving spin values after an anneal run. Since these are external factors (and even depend on the traffic on the machine) we have excluded them from our definition of quantum annealing machine-time. Indeed, these factors are the main obstruction preventing the reproduction of the saturation plot in Fig.~\ref{fig:sat_plot_2}.
Hence the dashed quantum annealer lines are built  upon the simulated annealing data by extrapolating comparative early performance. Specifically, the grey dashed line represents a projection of the quantum annealer performance, including the time required to perform the classical checks. 

As we can see from the figure, both quantum and simulated annealing are far more efficient than random scans and genetic algorithms (red and green solid lines, respectively). Modulo the above caveats, in this case we estimate that the quantum annealing method is $1.3$ times faster even than simulated annealing. The orange dashed line is an estimate of a hypothetical pure quantum annealing implementation, in which we assume that {\it all} of the additional classical checks are instead also encoded on the annealer (using methods that we discuss below), with no additional classical post-processing being required.  This would result in a $4.125$ times better performance than simulated annealing.

Overall then, in this study annealing methods appear to surpass other techniques in terms of performance, efficacy. They outshine alternative approaches and show superior outcomes. Quantitatively, the initial model discovery  rate (\ie, within the first $1000s$) is one model in $0.66$ seconds for simulated annealing, compared to one models in $33$ seconds for the genetic algorithm and one in $100$ seconds for the random scan. Using quantum annealing, the rate increases to one model in $0.50$ seconds when model checks are performed classically (the grey line in Fig.~\ref{fig:sat_plot_2}) and
to one model in $0.16$ seconds for a hypothetical purely quantum annealing implementation (orange dashed line). However one caveat we should add for completeness is that the search in this study may be somewhat disadvantageous for the GA as the number of SM-like models in the search space is relatively dense: as discussed in Ref.~\cite{Abel:2014xta} a GA becomes more effective when the search is very difficult, say one model in $10^{6}$ or more. The future challenge then is to compare quantum annealers and GAs when they are both confronted by much harder problems such as those discussed in Refs.~\cite{Abel:2021rrj,Abel:2023zwg}.

\section{Towards pure QA model building}

\label{sec:reduction}

In Fig.~\ref{fig:sat_plot_2} we included a 
line for hypothetical purely quantum annealing in which no classical post-processing would need to be done. In this section we briefly consider how such a complete implementation might be achieved. This discussion will illustrate the potential and also the complexity of performing pure quantum computing analyses. 

The first hurdle for a complete quantum implementation of the models under discussion is the fact that the only spin Hamiltonians that can currently be considered are quadratic Ising models. Let us suppose for this discussion that this will remain the case for the foreseeable future. (We shall comment on alternative possibilities at the end of this section.) The problem then is how do we encode the higher order constraints on the annealer. For example if we do not fix the values of $U_s^3$, $U_v^1$, $U_v^2$ in advance  but allow all the $U$ parameters to be set by the quantum annealing (in which case we are searching the larger  space of size $10^{12}$) we are then obliged to encode the squares of Eqs.~\eqref{eq:U_choice}
in the Hamiltonian, which would be quartic in the spins.

Thus we will need to know how to turn a higher order polynomial of spins into a quadratic Ising Hamiltonian on the annealer. This is done by the method of reduction, described in Refs.~\cite{Abel:2022lqr,Abel:2022wnt}. 
This method involves introducing auxilliary spins to represent pairs of spins in the original Hamiltonian, and it works as follows. 

Let us begin with the raw high order polynomial $\tilde{H}(\sigma_\ell)$ written as a function of binary QUBO variables using Eq.~\eqref{eq:taufromsig}. Suppose $\tilde{H}$ has terms involving products of two binary variables $\tau_1$ and $\tau_2$.
Now consider adding to the polynomial $\tilde{H}$ a quadratic term that involves the binary variables together with a new auxiliary variable $\tau_{12}$, which is of the form 
\begin{equation}
Q(\mbox{\small $\tau_{12};\tau_1,\tau_2$}) ~ = ~ \Lambda (\tau_1 \tau_2 - 2 \tau_{12} (\tau_1 + \tau_2) + 3 \tau_{12})~.
 \label{eq:constraint-Hamiltonian}
\end{equation}
Inspection shows that a sufficiently large and positive overall coupling $\Lambda$ enforces $\tau_{12} ~=~\tau_1\tau_2$ whatever the values of $\tau_1$ and $\tau_2$ happen to be. Importantly the  minimum at this point has $Q=0$. Therefore we may replace the product $\tau_1\tau_2$ with $\tau_{12}$ wherever it appears within $\tilde{H}$, and the new Hamiltonian is guaranteed to have the same set of minima as the original $\tilde{H}$. This process can be iterated until one arrives at the desired problem-Hamiltonian which is quadratic in spins, 
and which is schematically of the form 
\begin{align}
H ~&=~ \tilde{H}(\mbox{\small$\tau_1,\tau_2,\ldots, \tau_{12},\tau_{13},\ldots ,\tau_{12,34},\tau_{12,56}\ldots $}) \nonumber \\
& + \sum_{i>j} Q(\mbox{\small$\tau_{ij};\tau_i,\tau_j$}) +\sum_{i<j,k<m}
Q(\mbox{\small$ \tau_{ij,km};\tau_{ij},\tau_{km}$})\nonumber \\
& \qquad ~+\ldots 
\end{align}
with the constraints imposed by the $Q$ terms ensuring that this quadratic Hamiltonian has the same minima as the original polynomial.  Although this procedure may seem laborious it can easily be automated to systematically reduce any higher order polynomial in spins to a quadratic, as in the explicit examples in Ref.~\cite{Abel:2022wnt}. 

Thus in principle all the remaining degrees of freedom may be put on to the annealer with the GGSO constraints being entirely enforced by the Hamiltonian, at the expense of using an additional ancillary qubit for every term in the Hamiltonian that requires reducing. 

There remains the problem of incorporating phenomenological counting constraints. That is, suppose that the annealer with this larger system  has found a solution to Eqs.~\eqref{eq:system_annealer} having thereby determined a set of $U$'s and $\Delta$'s. We are required to include something in the Hamiltonian which will now implement the three generations check in Eq.~\eqref{eq:generations_constraints} which previously we did classically. This requires the introduction of ``counting'' qubits and it can be achieved as follows. 

We are required to impose Eq.~\eqref{eq:generations_constraints} which boils down to imposing $\# (\chi ={\rm even}) = 3+\# (\chi = {\rm odd}  )$, or in other words
\be 
\sum_{I=1}^3 \sum_{p,q,r,s \in \Xi_s^I} \left( {2~\overline\chi_{pqrs}-1}\right)  ~=~  - ~3~, 
\label{eq:3gconst}
\ee
where $\overline \chi \equiv  \chi~~{\rm mod}(2)$. Note that the quantity on the left is actually just given by the sum of the $\sigma$ qubits corresponding to the binary $\overline \chi$ variable. This is sufficient to ensure that even if we have cancelling positive and negative chiralities the nett number of generations is 3. 

In these equations the $\chi_{pqrs}$ are functions of the $p,q,r,s$ as in Eqs.\eqref{eq:chi^1},\eqref{eq:chi^2},\eqref{eq:chi^3}, so they are also polynomial objects in spins. However each $\chi$ can be mapped to a single ancillary binary qubit $\overline \chi$ by adding the terms 
\be
H~\supset~ ( \chi + 2 K_\chi -  \overline{\chi}  )^2
\label{eq:reduc1}
\ee
for every $\chi$, where here $\overline \chi$ is another ancillary binary qubit  and $K_\chi$ is as before a binary encoded integer, $K_\chi \in \mathbb Z$. Thus $\overline \chi = \chi$ is enforced at the minimum. Of course given that the $\chi_{pqrs}$ in 
Eqs.~\eqref{eq:chi^1},\eqref{eq:chi^2},\eqref{eq:chi^3}
are not linear in spins, Eq.~\eqref{eq:reduc1} will also require reduction. 
%\be
%H = Q(\tau_{\chi_r \zeta};  \chi_r , \zeta) + \Lambda' \tau_{\chi \zeta}^2 + (\chi_r -1)^2 + (\zeta-1)^2
%\ee 
Finally to impose the three generation constraint in Eq.~\eqref{eq:3gconst} we then add to the quadratic Hamiltonian the term
\be 
H~\supset ~ 
\left( 
\sum_{I=1}^3 \sum_{p,q,r,s \in \Xi_s^I} \left( {2 \overline\chi_{pqrs}-1}\right) + 3
\right)^2~
\ee
which is quadratic and therefore requires no further reduction.

In principle therefore all the consistency conditions may be straightforwardly implemented on the annealer in this fashion. Currently the limiting factor is the size of the architecture and the relatively large number of ancillary qubits that would be generated by this method. Therefore one might also contemplate an alternative approach which is to implement annealing on a quantum gate computer. In such an approach the Hamiltonian is ``Trotterized'' in order to  evolve the system in small time-steps on a universal gate quantum computer \cite{qibo_paper}. Thus in principle spin Hamiltonians of high order are allowed and no reduction would be required in order to implement all the consistency conditions. Currently no such system of large enough physical size is available to encode the system under discussion, and unfortunately one cannot simulate more than approximately 30 qubits. However this approach would be a promising avenue to explore once universal gate machines of sufficient size become available.  

\section{\label{Sec:Conclusions}Conclusions}

In this study, we have employed quantum annealing to construct string models, focusing on their efficiency and effectiveness in the model discovery process. By comparing quantum annealing with other established methods such as simulated annealing, random scans, and genetic algorithms, we have gained valuable insights into the possible advantages of using quantum annealers for this purpose.

We should add that annealers are possibly most advantageous when the search space consists of relatively dense regions of SM-like models (in this study one model in $10^5$), a situation in which  genetic algorithms do not usually lead to significant improvement with respect to alternative methods such as random scans. By contrast, genetic algorithms are known to excel in scenarios with more challenging searches, where the exploration of extremely large solution spaces is required. Therefore, it would be interesting in future investigations to compare these methods in more difficult problem domains in order to provide a comprehensive assessment of their respective strengths and weaknesses. \\

\noindent {\it {Acknowledgements}:}  We would like to thank Andrei Constantin, Thomas Harvey and Andrei Lukas for helpful discussions. We thank Victoria Goliber (D-wave) and Matteo Robbiati (Qibo) for disucssion and technical assistance. S.A. is supported by the STFC under grant ST/P001246/1. 

\appendix
\section{Construction of $SO(10)$ models}
\label{app1}
In this article we focus on heterotic string models defined in the free fermionic formulation using the
basis $b=\{\beta_1,\dots,\beta_{12}\}$,
where 
\begin{align}
\beta_1&=\mathds{1}=\{\psi^\mu,x^{1,\dots,6},y^{1,\dots,6},\omega^{1,\dots,6};
\nonumber\\
&\hspace{0.7cm}\bar{y}^{1,\dots,6},\bar{\omega}^{1,\dots,6},\bar{\psi}^{1,\dots,5},
\bar{\eta}^{1,2,3},\bar{\phi}^{1,\dots,4},\bar{\phi}^{5,\dots,8}
\}\,,\nonumber\\
\beta_2&=S=\{\psi^\mu,x^{1,\dots,6}\}\,,\nonumber\\
\beta_{2+i}&=e_i=\{y^i\omega^i;\bar{y}^{i},\bar{\omega}^{i}\}\,,i=1,\dots,6\,,
\\
\beta_{9}&=b_1=\{x^{34},x^{56},y^{3,4},y^{5,6};\bar{y}^{3,4},\bar{y}^{5,6},\bar{\psi}^{1,\dots,5},\bar{\eta}^{1}\}\,,\nonumber\\
\beta_{10}&=b_2=\{x^{12},x^{56},y^{1,2},y^{5,6};\bar{y}^{1,2},\bar{y}^{5,6},\bar{\psi}^{1,\dots,5},\bar{\eta}^{2}\}\,,\nonumber\\
\beta_{11}&=z_1=\{\bar{\phi}^{1,2,3,4}\}\,,\nonumber\\
\beta_{12}&=z_2=\{\bar{\phi}^{5,6,7,8}\}\,,\nonumber
\end{align}
and a set of phases $\cc{\beta_1}{\beta_1}=\pm1,\cc{\beta_i}{\beta_j}=\pm1,i>j=1,\dots,6$. 
The basis vectors $\beta_i$ describe the parallel transportation properties of the fermionic coordinates along the world-sheet torus while the phases link to generalised GSO projections (GGSO). Following the standard notation, included fermions are periodic, while all rest are anti-periodic. For $\cc{S}{e_i}=\cc{s}{z_a}=-1, i=1,\dots,6, a=1,2$ and generic choice of the remaining GGSO phases, the above basis describes 
a $\mathcal{N}=1$ supersymmetric model possessing  $SO(10){\times}U(1)^3{\times}SO(8)^2$ gauge symmetry.

$SO(10)$ spinorials arise from the sectors $\mathcal{S}^I_{\vec{P}^I_s}= S+b_I+\vec{P}^I_s{\cdot}\vec{E},\,I=1,2,3$  where $P^1_s=(0,0,p^1_s,q^1_s,r^1_s,s^1_s)$, $P^2_s=(p^2_s,q^2_s,0,0,r^2_s,s^2_s)$, $P^3_s=(p^3_s,q^3_s,r^3_s,s^3_s,0,0)$ and $\vec{E}= (e_1,e_2,e_3,e_4,e_5,e_6)$. Here,  $b^3=b^1+b^2+x$ with $x=\mathds{1}+S+\sum_{i=1}^6 e_i+\sum_{a=1}^2 z_a$.
Similarly, $SO(10)$ vectorials come from the sectors $\mathcal{V}^I_{\vec{P}^I_v}= S+b_I+x+\vec{P}^I_v{\cdot}\vec{E},\,I=1,2,3$  where $P^1_v=(0,0,p^1_v,q^1_v,r^1_v,s^1_v)$, $P^2_s=(p^2_v,q^2_v,0,0,r^2_v,s^2_v)$, $P^3_v=(p^3_v,q^3_v,r^3_v,s^3_v,0,0)$.

Utilising $e_i\cap \mathcal{S}^1_{\vec{P}^1_s}=\emptyset, i=1,2\,$,
$e_i\cap \mathcal{S}^2_{\vec{P}^1_s}=\emptyset, i=3,4\,$,
$e_i\cap \mathcal{S}^3_{\vec{P}^1_s}=\emptyset, i=5,6\,$ and
$z_a\cap \mathcal{S}^I_{\vec{P}^1_s}=\emptyset, a=1,2,\, I=1,2,3\,$
the spinorial projectors can be recast in the form 
\begin{align}
\Delta^I U_s^I=Y_s^I\ ,\ I=1,2,3\,,
\label{spin_eq}
\end{align}
where
\begin{align}
\Delta^1=
\left(
\begin{array}{cccc}
(e_1|e_3)&(e_1|e_4)&(e_1|e_5)&(e_1|e_6)\\
(e_2|e_3)&(e_2|e_4)&(e_2|e_5)&(e_2|e_6)\\
(z_1|e_3)&(z_1|e_4)&(z_1|e_5)&(z_1|e_6)\\
(z_2|e_3)&(z_2|e_4)&(z_2|e_5)&(z_2|e_6)\\
\end{array}
\label{eq:delta_1}
\right)\,,
\\\Delta^2=
\left(
\begin{array}{cccc}
(e_3|e_1)&(e_3|e_2)&(e_3|e_5)& (e_3|e_6)\\
(e_4|e_1)&(e_4|e_2)&(e_4|e_5)&(e_4|e_6)\\
(z_1|e_1)&(z_1|e_2)&(z_1|e_5)&(z_1|e_6)\\
(z_2|e_1)&(z_2|e_2)&(z_2|e_5)&(z_2|e_6)\\
\end{array}
\label{eq:delta_2}
\right)\,,
\\\Delta^3=
\left(
\begin{array}{cccc}
(e_5|e_1)&(e_5|e_2)&(e_5|e_3)&(e_5|e_4)\\
 (e_6|e_1)&  (e_6|e_2)&  (e_6|e_3)& (e_6|e_4)\\
(z_1|e_1)&(z_1|e_2)&(z_1|e_3)& (z_1|e_4)\\
(z_2|e_1)&(z_2|e_2)&(z_2|e_3)&(z_2|e_4)\\
\end{array}
\label{eq:delta_3}
\right)\,,
\end{align}
and
\begin{align}
U^1_s=
\left(
\begin{array}{c}
p^1_s\\
q^1_s\\
r^1_s\\
s^1_s
\end{array}
\right)
\ ,\ 
U^2_s=
\left(
\begin{array}{c}
p^2_s\\
q^2_s\\
r^2_s\\
s^2_s
\end{array}
\right)
\ ,\ 
U^3_s=
\left(
\begin{array}{c}
p^3_s\\
q^3_s\\
r^3_s\\
s^3_s
\end{array}
\right)\,
\end{align}
and
\begin{align}
Y^1_s=
\left(
\begin{array}{c}
(e_1|b_1)\\
(e_2|b_1)\\
(z_1|b_1)\\
(z_2|b_1)\\
\end{array}
\right)
\,,\,
Y^2_s=
\left(
\begin{array}{c}
(e_3|b_2)\\
(e_4|b_2)\\
(z_1|b_2)\\
(z_2|b_2)\\
\end{array}
\right)
\,,\,
Y^3_s=
\left(
\begin{array}{c}
(e_5|b_3)\\
(e_6|b_3)\\
(z_1|b_3)\\
(z_2|b_3)\\
\end{array}
\right)
\label{eq:Y_s}
\end{align}
For the surviving spinorials we calculate chiralities using the formulae
\cite{Faraggi:2004rq} , $X^{(I)}_{pqrs}=\exp(i \pi \chi^{(I)}_{pqrs})$, with
\begin{align}
\chi^{(1)}_{pqrs}&=\alpha_0+(1-r)(e_5|b_1)+(1-s) (e_6|b_1)\nonumber\\
&\ + p (e_3|b_2) + q (e_4|b_2) + r (e_5|b_2) + s (e_6|b_2)\nonumber\\
&\ + p(1-r) (e_3|e_5) + p(1-s) (e_3|e_6) \nonumber\\
&\ + q(1-r) (e_4|e_5) + q(1-s)(e_4|e_6) \nonumber\\
&\ +(r+s) (e_5|e_6)\,, 
\label{eq:chi^1}
\\
\chi^{(2)}_{pqrs}&=\alpha_0+(1-r)(e_5|b_2)+(1-s) (e_6|b_2)\nonumber\\
&\ + p (e_1|b_1) + q (e_2|b_1) + r (e_5|b_1) + s (e_6|b_1)\nonumber\\
&\ + p(1-r) (e_1|e_5) + q(1-r) (e_2|e_5) \nonumber\\
&\ + p(1-s) (e_1|e_6) + q(1-s) (e_2|e_6) \nonumber\\
&\ +(r+s) (e_5|e_6)\,, 
\label{eq:chi^2}
\\
\chi^{(3)}_{pqrs}&=\alpha_0 + (1-p) (e_1|b_1) + (1-q) (e_2|b_1)  \nonumber\\
&\  + (1-r) (e_3|b_2) + (1-s) (e_4|b_2) \nonumber\\
&\ + (1-r) (1-p) (e_1|e_3) + (1-r)(1-q) (e_2|e_3) \nonumber\\
&\ + (1-s) (1-p) (e_1|e_4) + (1-s) (1-q) (e_2|e_4) \nonumber\\
&\ + (1-r)\left[(e_3|e_5)+(e_3|e_6)\right]
 + (1-s)\left[(e_4|e_5)+(e_4|e_6)\right]\nonumber\\
 &\ + (1-r)\left[(e_3|z_1)+(e_3|z_2)\right]
 + (1-s)\left[(e_4|z_1)+(e_4|z_2)\right]\nonumber\\
&\ + (e_5|b_1)  + (e_6|b_1) + (z_1|b_1) + (z_2|b_1)\,,
\label{eq:chi^3}
\end{align}
where we can set $\alpha_0=0$ as it depends on conventions, 
$e^{i\pi\alpha_0} = - \text{ch}(\psi^\mu) c\left[\mathds{1} \atop S\right] c\left[S\atop b_1\right]c\left[S \atop b_2\right]c\left[b_1\atop b_2\right]$.

Similarly, vectorial projectors can be recast in the form
\begin{align}
\Delta^I U_v^I=Y_v^I\ ,\ I=1,2,3\,,
\label{vect_eq}
\end{align}
 where
\begin{align}
U^1_v=
\left(
\begin{array}{c}
p^1_v\\
q^1_v\\
r^1_v\\
s^1_v
\end{array}
\right)
\ ,\ 
U^2_v=
\left(
\begin{array}{c}
p^2_v\\
q^2_v\\
r^2_v\\
s^2_v
\end{array}
\right)
\ ,\ 
U^3_v=
\left(
\begin{array}{c}
p^3_v\\
q^3_v\\
r^3_v\\
s^3_v
\end{array}
\right)\,,
\end{align}
and
\begin{align}
Y^1_v=
\left(
\begin{array}{c}
(e_1|b_1^x)\\
(e_2|b_1^x)\\
(z_1|b_1^x)\\
(z_2|b_1^x)\\
\end{array}
\right)
\,,\,
Y^2_v=
\left(
\begin{array}{c}
(e_3|b_2^x)\\
(e_4|b_2^x)\\
(z_1|b_2^x)\\
(z_2|b_2^x)\\
\end{array}
\right)
\,,\,
Y^3_v=
\left(
\begin{array}{c}
(e_5|b_3^x)\\
(e_6|b_3^x)\\
(z_1|b_3^x)\\
(z_2|b_3^x)\\
\end{array}
\right)\,,
\label{eq:Y_v}
\end{align}
with $b_i^x=b_i+x$.
The GGSO associated coefficients in $Y^3_s, Y_v^I,i=1,2,3$ can be reduced 
as follows
\begin{align}
(e_i|b_3)&=(e_i|b_1)+(e_3|b_2)+(e_i|x)\\
(z_a|b_3)&=(z_a|b_1)+(z_a|b_2)+(z_a|x)\\
(e_i|b_I+x)&=(e_i|b_I)+(e_i|x)\,,I=1,2\\
(z_a|b_I+x)&=(z_a|b_I)+(z_a|x)\,,I=1,2\\
(e_i|b_3+x)&=(e_i|b_1)+(e_i|b_2)\,,\\
(z_a|b_3+x)&=(z_a|b_1)+(z_a|b_2)\,,
\end{align}
with
\begin{align}
(e_i|x)&= \sum_{j=1\atop j\ne i}^6(e_i|e_j) + (e_i|z_1)+(e_i|z_2)\,,\\
(z_a|x)&= 1+\sum_{j=1}^6(z_a|e_j) + (z_1|z_2)\,.
\end{align}
In the notation employed here
\begin{align}
c\left[\alpha\atop\beta\right]=\text{e}^{\text{i}\pi(\alpha|\beta)}\,.
\end{align}
Moreover,
\begin{align}
(e_i|e_j)=(e_j|e_i)\ ,\ (e_i|z_a)=(z_a|e_i)\ ,\\
(z_1|z_2)=(z_2|z_1)
\ ,\ (e_i|b_k)=(b_k|e_i)\,.
\end{align}
For a given set of spin structure coefficients $c\left[\beta_i\atop\beta_j\right]$ the solutions $U^I_s, U^I_v, I=1,2,3$ of \eqref{spin_eq}, \eqref{vect_eq} determine the 
number of surviving spinorials/vectorials. Without loss of generality, 
we can assume that one spinorial comes from the $S+b_1$ sector, that is
we have a solution with $p^1_s=q^1_s=r^1_s=s^1_s=0$. This amounts to
setting 
\begin{align}
(e_1|b_1)=(e_2|b_1)=(z_1|b_1)=(z_2|b_1)=0\,
\end{align}
as dictated by the relevant equation of \eqref{spin_eq}, i.e  $\Delta^1 U^1_s=Y^1_s$. Furthermore, we
can also assume that the second spinorial arises from $S+b_2$, i.e.  that $p^2_s=q^2_s=r^2_s=s^2_s=0$ is a solution of $\Delta^2 U^2_s=Y^2_s$, 
which then implies
\begin{align}
(e_3|b_2)=(e_4|b_2)=(z_1|b_2)=(z_2|b_2)=0\,.
\end{align}
The existence of a coupling of the form $\mathbf{16}\times\mathbf{16}\times\mathbf{10}$ at the trilinear effective superpotential (top mass Yukawa  coupling) requires at least one vectorial coming from $S+b_1+b_2+x$, that is Eq. \eqref{vect_eq}
has a solution  with $p^3_v=q^3_v=r^3_v=s^3_v=0$ \cite{Rizos:2014uba}. Consequently, we also set
\begin{align}
(e_5|b_1)=(e_5|b_2)\ ,\ (e_6|b_1)=(e_6|b_2)\ , \label{eq:yukawa_1}\\
(z_1|b_1)=(z_1|b_2)\ ,\ (z_2|b_1)=(z_2|b_2)\,
\label{eq:yukawa_2}
. 
\end{align}
Finally additional constraints come from adjusting  spinorial chiralities, in order to satisfy the chirality constraints on the first and second plane -- i.e. to solve the equivalent of Eq.~\eqref{eq:chi0}
for Eqs.~\eqref{eq:chi^1} and \eqref{eq:chi^2}. These conditions are
\begin{align}
\label{eq:a29}
(e_5|b_1)=(e_6|b_1)\ ,\ (e_5|b_2)=(e_6|b_2)\,. 
\end{align}

\bibliographystyle{inspire}
\bibliography{references,referencesSAMS}

\end{document}